\documentstyle[prl,twocolumn,aps]{revtex}

\begin{document}
 \tolerance 50000

\draft

\title{Equivalence of the
Variational Matrix Product Method and the 
Density Matrix Renormalization Group applied to Spin Chains} 
\author{J. Dukelsky$^{1}$,  
M.A. Mart\'{\i}n-Delgado$^{2}$,
T. Nishino$^{3}$ and 
G. Sierra$^{4}$
 } 
\address{ 
$^{1}$Instituto de Estructura de la Materia, C.S.I.C.,Madrid, Spain.
\\ 
$^{2}$Departamento de
F\'{\i}sica Te\'orica I, Universidad Complutense. Madrid, Spain.
 \\
$^{3}$Department of Physics, Faculty of Science, Kobe University,
Japan.\\
$^{4}$Instituo de Matem\'aticas y F\'{\i}sica Fundamental, C.S.I.C.,
Madrid, Spain. 
}

\twocolumn[
\maketitle 
\widetext

\vspace*{-1.0truecm}

\begin{abstract} 
\begin{center}
 \parbox{14cm}{We  present a rotationally invariant   
matrix product method (MPM) of isotropic 
spin chains. This allows us to deal with a larger  
number of variational MPM parameters than those considered 
earlier by other authors. We also show  
the  relation between the MPM and the DMRG method of White. 
In our approach the eigenstates of the density matrix associated 
with the MPM are used 
as variational
parameters together with the standard MPM parameters. We compute 
the ground state energy density and the 
spin correlation length of the spin 1 Heisenberg chain. }
\end{center}
\end{abstract}

\pacs{
 \hspace{2.5cm} 
}

]
\narrowtext

The density matrix renormalization group (DMRG), introduced by
White\cite{W} in
1992, is a powerful numerical method to  
study the ground state  and 
static properties of quantum lattice systems,
as for example the Heisenberg, t-J, and  Hubbard models defined on 
chains, ladders
and clusters. The DMRG uses the Wilsonian scheme of 
adding one point at each RG step. After many
iterations the DMRG reaches a fixed point, and 
the ground state exhibits a matrix product structure (MP),
as was shown by Ostlund and Rommer\cite{OR}. 
These authors proposed to start from a MP ansatz
for the ground state of the system, determining  
the variational parameters by
the standard variational method, without resorting  to the DMRG.
The advantage of the MPM is that it is analytical 
and does not require big computational resources. However
it is not clear how to treat large values of the number
of states $m$ used in the minimization of the energy.  
On the other hand it is not clear the relation 
between the MPM and the DMRG, apart from 
sharing a MP structure in the thermodynamic limit.

In this letter we show i) how to treat numerically
and analytically larger 
values of $m$ than those considered by other authors, 
thanks to the reduction of the 
basis obtained by exploiting 
the rotational symmetry of the problem and 
ii) exhibit the relation between the MPM
and  the DMRG. In particular we shall see that the MPM 
naturally leads to a density matrix whose eigenvalues
appear  as variational parameters together with those
that  generate the MP ansatz. 
In fact our formalism is closely connected to that 
developed in ref \cite{FNW} ( see also \cite{A,K}). 
We apply the MPM and the DMRG to the
spin 1 Heisenberg chain, and compare the results obtained 
with both methods.
For the ground state energy density, the MPM gives
a better estimate for all  values of $m >1$,
which we interpret as been caused by the existence of
a bound state in the middle of the superblock in the 
DMRG method. In the MPM this bound state is absent by 
construction.
For increasing values of $m$ the discrepancy between the MPM and
the DMRG tends to disappear. The numerical 
results for the eigenvalues of the density
matrices in both methods are similar, and they seem to
converge to a common value when increasing $m$.
 From these results we conclude the equivalence between the MPM
and the DMRG methods in the thermodynamic limit  
for large values of $m$.

We shall consider a spin chain with spin $S$ at each site,
where $S$ is an integer ( the case of half integer spin will 
be treated in a separate work).
Let us denote the basis states of the MPM as
$|a,J M\rangle_N$,   where $N$ is the length of the chain, 
$a = 1, \cdots , d_J$ denotes 
the multiplicity of the total spin $J$ of the state
and $M$ is the third component of the spin. Counting the  number of
multiplets we have $m= \sum_J \; d_J$, which 
correspond to a number $m_W = \sum_J (2 J +1) \; d_J$
of states in the standard DMRG.  
At the fixed point of the DMRG one has the matrix product
law\cite{OR},

\begin{eqnarray}
&\left| a_1,J_1 M_1\right\rangle_{N} =\sum_{a_2 J_2,M_2,M} \; A_{a_1J_1,a_2J_2}
& \label{1} 
\end{eqnarray}

\begin{eqnarray}
& \times 
\left| S M\right\rangle _{N}\ \otimes \left| a_2,J_2M_2\right\rangle_{N-1}\
\left\langle S M,J_2M_2\right. \left| J_1 M_1\right\rangle  &
\nonumber 
\end{eqnarray}

\noindent
where
$\left\langle S M,J_2M_2\right. \left|
J_1,M_1\right\rangle$ are Clebsch-Gordan coefficients and
$A_{a_1 J_1, a_2 J_2}$ are variational parameters
subject to the following conditions

\begin{eqnarray}
& A_{a_1 J_1, a_2 J_2} = 0 \;\; {\rm unless} \;\;
|J_2 - S| \leq J_1 \leq J_2  + S  & \label{2} \\
& \sum_{a_2,J_2}A^*_{a_1 J_1,a_2 J_2} \; A_{a'_1 J_1,a_2 J_2}= 
\delta_{a_1 a'_1} & 
\label{3} 
\end{eqnarray}

Eq.(\ref{2}) follows from the 
CG decomposition $ S \otimes J_2 \rightarrow J_1$ in (\ref{1}), while
condition (\ref{3}) guarantees that the states 
$|a, J M \rangle_N $ constitute an 
orthonormal basis for all values of $N$. 
The initial data of the recurrence relation (\ref{1}) is given by
choosing a spin $S/2$ irrep at the end of the chain.
This choice  eliminates the  
multiplicity associated to the 
effective spins $S/2$ at the ends of the chain.
The sum in $J's$ in eq. (\ref{1}) is of course restricted
to a finite set of spins.

The parameters $A_{a_1 J_1, a_2 J_2}$ are determined by 
minimizing the energy of the states $|a, J M \rangle_N$ in the limit
where $N \rightarrow \infty$. For this purpose let us define the
following quantity,

\begin{equation}
E^{(N)}_{a a^{\prime }J}=_{N}\left\langle a,JM\right| H_{N}\left| a^{\prime
},JM\right\rangle _{N} 
\label{4}
\end{equation}

\noindent where $H_N$ is the Hamiltonian acting on the chain with $N$
sites. From eq.(\ref{1}) one can derive a recursion formula for 
$E^{(N)}_{a a^{\prime }J}$,

\begin{equation}
E^{(N)}_{a a^{\prime }J} = V_{a a^{\prime }J}
+ \sum_{b b' J'} \; T_{a a' J, b b' J'} \;  
E^{(N-1)}_{b b^{\prime }J'} , \;\;\;\; ( N \geq 3)
\label{5}
\end{equation}

\noindent where ${\bf T}$ is  a matrix with 
entries ( we assume from now on the reality of $A_{a_1 J_1 , a_2 J_2}$),

\begin{equation}
T_{a a' J, b b' J'} = A_{a J, b J'} \; A_{a' J, b' J'}
\label{6}
\end{equation}

\noindent and  $ W$ is the matrix element of the 
piece of the Hamiltonian which couples the sites $N$ and $N-1$,
which does not dependent on $N$,

\begin{equation}
V_{a a^{\prime }J} = 
_{N}\left\langle a,JM\right| H_{N-1,N}\left|
a^{\prime },JM\right\rangle _{N} 
\label{7}
\end{equation}

For the Heisenberg model, $H_{N-1,N} = {\bf S}_{N-1} \cdot {\bf S}_N$, 
and applying 
the Wigner-Eckart theorem,  we find 
the following expression  for $W$ in terms of 6-j symbols,

\begin{eqnarray}
& V_{a_1 a_2 J_1}= 
\sum_{a_3J_2,a_4J_3,a_5J_4} 
\;  {\cal H}_{J_1 J_2 J_3 J_4} & \label{8} \\
& \; A_{a_1J_1,a_3J_2}
A_{a_2J_1,a_4J_3} \; A_{a_3J_2,a_5J_4}A_{a_4J_3,a_5J_4} ,\ &
\nonumber 
\end{eqnarray}

\begin{eqnarray}
&{\cal H}_{J_1,J_2,J_3,J_4}=\left( -\right) ^{2S+J_1+J_2+J_3+J_4+1}S\left(
S+1\right) \left( 2S+1\right) &\label{9} \\
&\times 
\sqrt{\left( 2J_2+1\right) \left(
2J_3+1\right) }
\left\{ 
\begin{array}{ccc}
1 & S & S \\ 
J_1 & J_2 & J_3
\end{array}
\right\} \left\{ 
\begin{array}{ccc}
1 & S & S \\ 
J_4 & J_2 & J_3
\end{array}
\right\}  & \nonumber
\end{eqnarray}

The solution of (\ref{5}) 
can be expressed in matrix notation as,

\begin{equation}
|{E}^{(N)} \rangle = \left( 1 + {\bf T} + {\bf T}^2 + \dots + 
{\bf T}^{N-3} \right)
\; |W \rangle  + {\bf T}^{N-2} \; |E^{(2)} \rangle 
\label{10}
\end{equation}

\noindent 
where $|{E}^{(N)}\rangle$ is regarded  in (\ref{10}) as  a vector
whose components are  labeled by $(a a' J)$. The entries of
${\bf T}$ are given by eq.(\ref{6}).

In the  limit $N \rightarrow \infty$ the
contribution from $|E^{(2)}\rangle$ drops off and 
we  shall show below  that $E^{(N)}_{a a^{\prime }J}$ behaves as,

\begin{equation}
\lim_{N \rightarrow \infty } \frac{1}{N} \;  E^{(N)}_{a a^{\prime }J} 
= \; \delta_{a a'} \;\; e_{\infty}
\label{11}
\end{equation}

\noindent 
where $e_\infty$ can be identified with the ground
state energy density and it  reads,

\begin{equation}
e_{\infty} = \sum_{a a' J} \; \rho_{a a' J} \; V_{a a^{\prime }J}
\label{12}
\end{equation}

In eq.(\ref{12}) $\rho_{a a' J}$ is the right eigenvector of
the matrix ${\bf T}$ with eigenvalue 1, and plays the role 
of a density matrix in the MPM. 
The proof of eqs.(\ref{11}) and (\ref{12}) follows from the
existence of an eigenvalue of
the matrix ${\bf T}$ equal to 1 \cite{FNW,OR}.
This property can be deduced from the normalization condition
(\ref{3}). Let us call $|{v}\rangle$ and $\langle{\rho}|$ the
right and left eigenvectors associated to the eigenvalue 1
of ${\bf T}$, which we shall assume to be  unique,

\begin{eqnarray}
& {\bf T} \; |{ v}\rangle = |{v}\rangle   &
\label{13} \\
& \langle{\rho}| \; {\bf T} = \langle \rho | & 
\label{14}
\end{eqnarray}

Then eq. (\ref{3}) implies that $|v \rangle$
is given in components by  $v_{a a' J } = \delta_{a a'}$..
On the other hand the quantities  
$\rho_{a a' J}$  that appear in (\ref{12}) are nothing but 
the components of
$\langle \rho|$, and
they are  found  by solving  the eigenvalue problem (\ref{14}). 
In eq.(\ref{12}) we have normalized
$\langle \rho|$  according to,

\begin{equation}
\langle \rho| v \rangle = 1 \; \rightarrow \; \sum_{a a' J} \;
\rho_{a a' J} = 1
\label{15}
\end{equation}

In a field theoretical language $|v\rangle$ and $\langle \rho|$
play the role  respectively of  incoming $|0 \rangle$ 
and outgoing vacua $ \langle 0|$, which are
left invariant  by the transfer matrix  operator ${\bf T}$,
that shifts by one lattice space the spin chain.
On the other hand 
$\rho_{a a' J}$, has the properties of a density 
matrix, and it corresponds precisely to  
the reduced density matrix of the blocks in the  
DMRG formalism, as we shall show below. 
It is remarkable  that the MP ansatz (\ref{1}) leads 
in a natural way  to 
a density matrix formalism.
This suggest that 
a rigorous mathematical formulation  
of the DMRG could perhaps be achieved within the
MPM.

By analogy with the DMRG we may choose a basis  where
the density matrix becomes diagonal, i.e. $\rho_{a a' J} =
w_{a J}^2 \; \delta_{a a'}$. Under this
assumption eq.(\ref{14}) reads,

\begin{equation}
\sum_{a_1 J_1} \; w^2_{a_1 J_1} \; A_{a_1 J_1 , a_2 J_2}
\; A_{a_1 J_1, a'_2 J_2} = \; w^2_{a_2 J_2} \; \delta_{a_2 a'_2}
\label{16}
\end{equation}

A solution of eqs.(\ref{16})  is obtained using 
eq.(\ref{3}) and assuming 
the following ``detailed balanced" condition \cite{NO},

\begin{equation}
w_{a_1 J_1} \; A_{a_1 J_1, a_2 J_2} \;
= \; w_{a_2 J_2} \;  A_{a_2 J_2, a_1 J_1}
\label{17}
\end{equation}

This eq. is very useful since we can eliminate almost a half of
the $A's$ in terms of the other half, and use
the $w's$ as independent variationally 
parameters. 
Hence the  problem reduces to  
the minimization of the GS energy (\ref{12}) with  
respect to the variational parameters $w_{a J}$ and 
$A_{a_1 J_1 , a_2 J_2}$ subject to the constraints 
(\ref{2}),(\ref{3}) and (\ref{17}).  For a MP ansatz with
no multiplicities, i.e. $d_J =1$, one can solve
all the constraints in terms of an 
independent set of parameters, however
when $d_J >1$ it is more efficient to use a numerical program
of minimization with constraints.

Taking into account all the variables and
constraints one sees that the  total number of independent 
variational parameters, $N_A$, 
is given by,

\begin{equation}
N_A = \frac{1}{2} \sum_{J_1 \neq J_2} \Lambda_{J_1 J_2} \; 
\; d_{J_1} \; d_{J_2} \;+ \sum_{J} \; d_J   -1
\label{18}
\end{equation}

\noindent where $\Lambda_{J_1 J_2}$ is 1 if $J_1$ and $J_2$
satisfy eq.(\ref{2}) and zero otherwise.

In table 1 we present the results for the case of spin $S=1$,
obtained with  the MPM and a version of the 
DMRG where the rotational symmetry has been 
used to eliminate the redundancy in the states kept\cite{SN}.
The case $m=1$ corresponds to the AKLT wave
function \cite{AKLT,AAH}, where $e_\infty$ computed with the
MPM and the DMRG coincide. This is because from eq.(\ref{18})
there is no adjustable parameter in the ansatz. The case
$m=4$ is the one considered in \cite{OR}.
The ground state energy $e_{\infty}(m)$ obtained with the MPM is always
lower than DMRG energy, this is related to the fact that the
wave function generated by  the infinite system DMRG is not uniform.
The DMRG optimizes the ground state of the renormalized system
[$B$]~$\bullet$~$\bullet$~[$B$], where [$B$] denotes the $m$-state
block spin, while  the super block [$B$~$\bullet$] has $3 m$ degrees of
freedom. From the view point of the MPM, the super block
[$B$~$\bullet$] should be optimized with $m$ degrees of
freedom. As a result, a shallow bound-state appears between
left-half of the system [$B$]~$\bullet$ and the right half
$\bullet$~[$B$], and the numerical precision in the ground state energy
is spoiled in DMRG, especially when $m$ is small. A way to improve the
DMRG from this error is to consider a system [$B$]~$\bullet$~[$B$]
at the last several steps in the infinite/finite system
DMRG algorithm. By choosing the block configuration, the degree of
freedom of the superblock [$B$~$\bullet$] is automatically
restricted to $m$ because the `reservoir' [$B$] has only $m$
degrees of freedom.

In order to compute the 
spin-spin correlation lengths $\xi$  of the MP states (\ref{1}),
we have to find out a recursion formula for the reduced matrix elements
of the spin operators ${\bf S}$, which is given by

\begin{eqnarray}
&_N \langle  a_1 J_1 || {\bf S}_1 || a_2 J_2 \rangle_N  &
\label{19} \\
& =\sum_{a_3 J_3 a_4 J_4} \; T^{(1)}_{a_1 J_{1} a_2 J_{2}, a_3 J_{3} a_4 J_4}
\; _{N-1} \langle  a_3 J_3 || {\bf S}_1 || a_4 J_4 \rangle_{N-1} & 
\nonumber
\end{eqnarray}

\noindent with

\begin{eqnarray}
& T^{(1)}_{a_1 J_{1} a_2 J_{2}, a_3 J_{3} a_4 J_4}=\left( -\right)
^{J_2+J_3+S+1}A_{a_1J_1, a_3J_3}A_{a_2J_2,a_4J_4} & \label{20} \\
& \times \sqrt{\left( 2J_1+1\right)
\left( 2J_2+1\right) } \;
\left\{ 
\begin{array}{ccc}
J_3 & J_1 & S \\ 
J_2 & J_4 & 1
\end{array}
\right\}  & \nonumber 
\end{eqnarray}

The correlation length $\xi$ is then given by the  highest 
eigenvalue $\lambda$, in 
absolute value,  of ${\bf T^{(1)} }$ by the formula,

\begin{equation}
\xi= - 1/{\rm ln} \lambda
\label{21}
\end{equation}

For the AKLT case (i.e. $d_{J}= \delta_{J, 1/2}$), eqs.(\ref{19}) 
and (\ref{20}) reproduce the exact spin-spin correlator found
in ref.\cite{AAH}. 
To analyze in more detail 
the relation MPM versus DMRG we give  in table 2
the eigenvalues of the matrix $\rho_{a a' J}$,
and those of the DMRG reduced density matrix in the case where
$m=6$. 
The later matrix 
has dimension $3 m$ and the  truncation DMRG method
consists in choosing  $m$ 
states with highest eigenvalues $w^2_{\rm DMRG}$, 
which add up to $1-P_m$ (see table 1).
For this reason we have to scale the DMRG weights 
of the states kept  
in order that they sum up to 1.

\begin{equation}
\overline{w}^2_{\rm DMRG} = w^2_{\rm DMRG}/ P_m
\label{22}
\end{equation}

In summary the results shown in  tables 1 and 2 suggest
that the predictions made by  
the MPM and the DMRG should become identical for large values of $m$.
In a later publication we shall present 
the results for the spin 
gap and other observables for various spin chains and ladders
using the MPM. An interesting problem is the  generalization of 
the MPM to the case of holes. The results of reference \cite{SMDWS} 
concerning the tJ ladder suggest that this generalization is possible
and worthwhile studying it.

{\bf Acknowledgements}: We would like to thank
J.M. Roman  for useful conversations. 
TN thanks Dr. Okunishi and Dr. Hieida for valuable discussions.
Numerical calculations were done by the SX-4 at the computer center of
Osaka university. JD acknowledges support from the DIGICYT under
contract No. PB95/0123, MAMD and GS acknowledges support from
the DIGICYT under contract No. PB96/0906.

\newpage

\begin{center}
\begin{tabular}{|c|c|c|c|c|c|c|c|c|}
\hline
$m$ & $N_A$ & $d_{1/2}$ & $d_{3/2}$ & $d_{5/2}$ & $-e_\infty ^{\rm MP}$ & $%
-e_\infty ^{\rm DMRG}$ & $1-P_m$ & $\xi ^{\rm MP}$ \\ 
\hline
\hline
$1$ & $0$ & $1$ & $0$ & $0$ & $1.333333$ & $1.333333$ &  
$1.6\times 10^{-2}$ & 0.910  \\ 
\hline
$2$ & $2$ & $1$ & $1$ & $0$ & $1.399659$ & $1.369077$ & $1.4\times 10^{-3}$
&2.600  \\ 
\hline
$3$ & $4$ & $2$ & $1$ & $0$ & $1.401093$ & $1.392515$ & $1.3\times 10^{-5}$
&3.338  \\ 
\hline
$4$ & $7$ & $2$ & $2$ & $0$ & $1.401380$ & $1.401380$ & $1.6\times 10^{-5}$
& 3.937 \\ 
\hline
$5$ & $10$ & $2$ & $2$ & $1$ & $1.401443$ & $1.401436$ & $7.6\times 10^{-6}$
& 4.085  \\ 
\hline
$6$ & $13$ & $2$ & $3$ & $1$ & $1.401474$ & $1.401468$ & $1.3\times 10^{-6}$
& 4.453\\
\hline
\end{tabular}
\end{center}
\begin{center}
Table 1. $m$ is total the number of multiplets, $N_A$ is the number
of independent variational parameters, $d_J$ is the number of multiplets
with spin $J$, $e_{\infty}^{{\rm MP},{\rm DMRG}}$ is 
the GS energy density of the MPM (DMRG), $1-P_m$ is the
probability of the states truncated out in the DMRG 
and $\xi^{\rm MP}$ is the spin correlation 
length of the MP state. The exact results
are given by $e_{\infty} = 1.4014845$ and $\xi= 6.03$ \cite{WH}.
\end{center}

\vspace{2cm}
\begin{center}
\begin{tabular}{|c|c|c|c|}
\hline
$a$  & $J$ & $w_{MP}^2$ & $\overline{w}_{DMRG}^2$ \\ 
\hline
\hline
$1$ & $1/2$ & $.9695581$ & $.9696232$ \\ 
\hline
$2$ & $1/2$ & $.0007662$ & $.0007599$ \\ 
\hline
$1$ & $3/2$ & $.0295443$ & $.0294877$ \\ 
\hline
$2$ & $3/2$ & $.0001119$ & $.0001089$ \\ 
\hline
$3$ & $3/2$ & $.0000078$ & $.0000085$ \\ 
\hline
$1$ & $5/2$ & $.0000118$ & $.0000118$ \\
\hline
\end{tabular}
\end{center}
\begin{center}
Table 2. $a$ and $J$ are the labels of the irrep,  
$w_{MP}^2$, are the eigenvalues of the MP density matrix,
and  $\overline{w}_{DMRG}^2$ are the corresponding DMRG
eigenvalues kept in the RG process
and normalized to 1. The data correspond to $m=6$. 
\end{center}

\end{document}